\documentclass[twocolumn,showpacs,preprintnumbers,amsmath,amssymb]{revtex4}

\usepackage{graphicx}
\usepackage{dcolumn}
\usepackage{bm}
\usepackage{color}
\usepackage{float}

\begin{document}


\title{Exploring contributions from incomplete fusion in $^{6,7}$Li+$^{209}$Bi and $^{6,7}$Li+$^{198}$Pt reactions}

\author{V. V. Parkar$^{1}$\footnote{vparkar@barc.gov.in}}
\author{V. Jha$^1$\footnote{vjha@barc.gov.in}}
\author{S. Kailas$^{1,2}$}

\affiliation{$^1$Nuclear Physics Division, Bhabha Atomic Research Centre, Mumbai - 400085, India}
\affiliation{$^2$UM-DAE Centre for Excellence in Basic Sciences, Mumbai - 400098, India}

\begin{abstract}
We use the breakup absorption model to simultaneously describe the measured cross-sections of the Complete fusion (CF), Incomplete fusion (ICF), and Total fusion (TF) in nuclear reactions induced by weakly bound nuclei $^{6,7}$Li on $^{209}$Bi and $^{198}$Pt targets. The absorption cross-sections are calculated  using the Continuum Discretized Coupled Channels (CDCC) method with different choices of short range imaginary potentials to get the  ICF, CF and TF cross-sections. It is observed that the cross-sections for deuteron-ICF/deuteron-capture are of similar magnitude as the $\alpha$-ICF/$\alpha$-capture, in case of $^{6}$Li projectile, while the cross-sections for triton-ICF/triton-capture is more dominant than $\alpha$-ICF/$\alpha$-capture in case of $^{7}$Li projectile. Both these observations are also corroborated by the experimental data. The ratio of ICF to TF cross-sections, which defines the value of fusion suppression factor is found to be in agreement with the data available from the literature. The cross-section ratio of CF/TF and ICF/TF show opposite behavior, the former decreases while the latter increases as the energy is lowered, which shows the dominance of ICF at below barrier energies. We have also studied the correlation of the ICF cross-sections with the non-capture breakup (NCBU) cross-sections as a function of energy, which shows that the NCBU is more significant than ICF at below barrier energies.

\end{abstract}

\maketitle

\section{\label{sec:Intro} Introduction}
In recent years, a large number of measurements of the fusion cross-sections involving weakly bound nuclei, such as, $^{6,7}$Li and $^{9}$Be, on several targets have been performed at energies around the Coulomb barrier \cite{Canto15}. Due to the low breakup threshold of the projectile nucleus, the breakup process has a significant role in description of fusion phenomenon in these reactions. In many of these studies, the measured fusion cross-sections are found to be significantly suppressed with respect to the coupled channel calculations at the above barrier energies \cite{Dasgupta04,Mukherjee06,Gas09,Rath09,Parkar10,Harphool12,Palshetkar10,Pradhan11,Rath13,Wu03,Gomes06,Hu15}. Similar coupled channel calculations for fusion cross-sections in  strongly bound nuclei, which account for the  couplings of the inelastic and transfer processes, successfully explain the measured fusion cross-sections. The suppression that is observed in reactions with weakly bound nuclei, is almost a universal phenomena observed in a large number of systems involving weakly bound projectiles for varying atomic masses of the  target nucleus \cite{Parkar10,Harphool12}. The suppression phenomena is largely attributed to the loss of flux subsequent to the breakup of the weakly bound projectile. There is a large probability that one or more fragments emanating from the breakup process may subsequently fuse with the target nucleus to form a compound nucleus. Therefore, for a consistent explanation of the fusion suppression, the fusion cross-section is often divided into two categories, namely, the complete fusion (CF) and the incomplete fusion (ICF). In the CF process, the whole projectile or all its charged fragments fuse with the target nucleus, whereas only a part of projectile is captured by the target in the ICF process and the remaining part escapes. These two processes together give a measure of the total fusion (TF) process. The knowledge of the relative contribution of the ICF and the CF is essential for understanding the suppression phenomena at energies above the barrier.

Despite considerable efforts in last several decades, quantum mechanical calculations of individual CF and ICF cross-sections are very scarce in the literature. In the CDCC based coupled channel calculations, which is a fully quantum mechanical treatment,  Hagino $\textit{et al.}$ \cite{Hagino00} and Diaz-Torres and Thompson \cite{Diaz02} performed calculations by classifying `complete fusion' as absorption from ground state, and `incomplete fusion' as absorption from breakup states. However, this method does not provide suitable description for the  weakly bound nuclei that dissociate into fragments of comparable masses, because the absorption of centre of mass of the projectile modeled in these works is not necessarily connected to the capture of all the fragments \cite{Diaz03}. In another work using CDCC calculations by Rusek $\textit{et al.}$ \cite{Rusek04}, a good description of certain observables such as, the total fusion (i.e., the sum of ICF and CF), the elastic, and non-capture breakup (NCBU) cross-sections was obtained. In very recent times, models based on post-form theory for the calculation of the inclusive breakup cross-sections has been successfully applied for $^{6}$Li+$^{209}$Bi system \cite{Lei15}. Apart from these efforts, approximate methods such as, a model based on a classical trajectories, which parameterizes ICF in terms of breakup probabilities  for weakly bound nuclei at energies around the Coulomb barrier have been developed by Diaz-Torres \cite{Diaz-Torres11}. A recent model by Boselli and Diaz-Torres \cite{Boseli15}, uses time-dependent wave-packet perspective for separating CF and ICF processes. In another work by Marta $\it{et~al.}$ \cite{Marta2014}, the semiclassical model calculations were attempted to understand CF and ICF data for $^{6,7}$Li+$^{209}$Bi, which however, did not describe the data satisfactorily. In many of these works, the absorption from the breakup channel is a crucial ingredient for calculation of ICF. The couplings arising from the excitation to the projectile continuum using the CDCC approach has provided a successful method for modeling the breakup process. Since the breakup process and the ICF arising due to the absorption following the breakup are intricately related, it is important to understand how breakup process is correlated with the ICF components.

For a comprehensive understanding of the fusion process with weakly bound nuclei, the relative contribution of the ICF and CF as a function of energy around the Coulomb barrier needs to be investigated well. In our earlier work, we attempted to obtain the use of absorption of the projectile fragments within the CDCC approach, which could provide a reasonable simultaneous explanation of CF and TF data with $^{9}$Be projectile over a large energy and target mass range \cite{9BePRC}. The absorption cross-sections obtained from the CDCC calculations were found to also successfully explain the universal suppression of fusion for the $^{9}$Be projectile with different targets. In these calculations, absorption cross-sections obtained using only one part of the fragment target potential is used to calculate the ICF. The $^{6,7}$Li, with their well defined cluster structure are quite amenable to this approach for calculations of ICF, CF and TF simultaneously. However, there is only a few data available for the simultaneous measurement of CF, ICF and TF for comparison with the calculations and most of the data available is either the CF or TF data.  In the present work, we have carried out calculations with the breakup-absorption model for the  $^{6,7}$Li+$^{209}$Bi and $^{6,7}$Li+$^{198}$Pt systems. For the  $^{6,7}$Li+$^{209}$Bi \cite{Dasgupta04} systems, the cross-section data of CF, ICF and TF is available over a large energy range. In addition, limited ICF data along with the CF data are available for the $^{6,7}$Li+$^{198}$Pt \cite{Shrivastava09,Shrivastava13} systems. These data have been utilized for the present investigations. Using the data and the calculations, we attempt to study the relative importance of ICF fraction in the fusion process and also investigate how the ICF depends on the non-capture breakup as a function of energy.

\section{\label{sec:Caln} Calculation Details}
\begin{figure}
\includegraphics[width=0.44\textwidth,trim=0.3cm 7.5cm 10.5cm 1cm, clip=true]{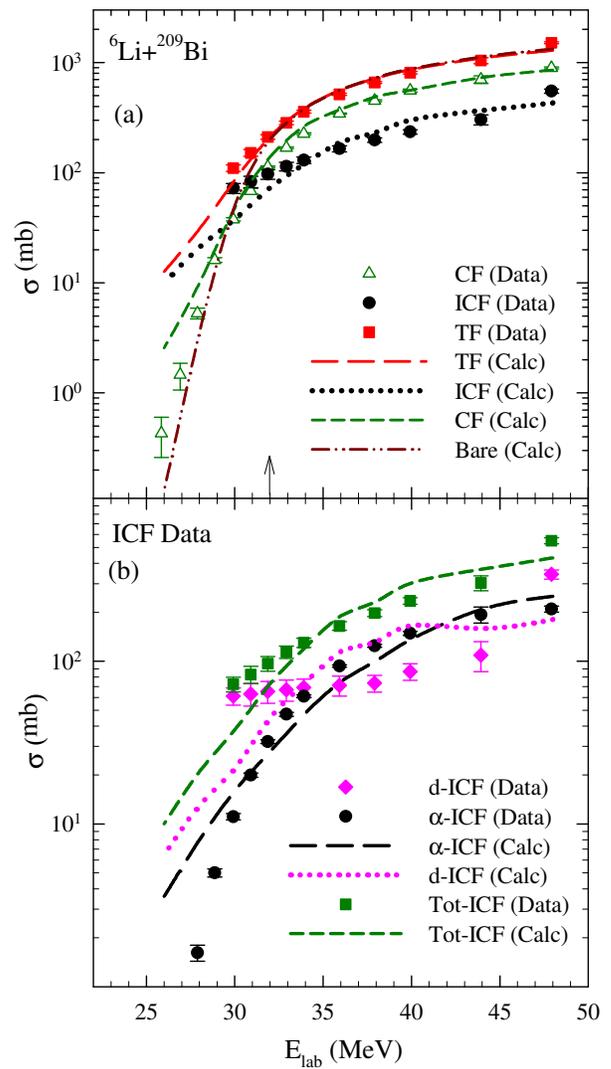}
\caption{\label{fig1}(Color online)(a) The data of Complete Fusion (CF), Incomplete Fusion (ICF) and Total fusion (TF)=CF+ICF+Fission for $^{6}$Li+$^{209}$Bi reaction from Ref.\ \cite{Dasgupta04} is compared with the calculations. The arrow indicate the position of Coulomb barrier. (b) Comparison of individual ICF contributions from $\alpha$-ICF, d-ICF along with Tot-ICF with the calculations. (see text for details).}
\end{figure}

\begin{figure}
\includegraphics[width=0.44\textwidth,trim=0.3cm 7.5cm 10.5cm 1cm, clip=true]{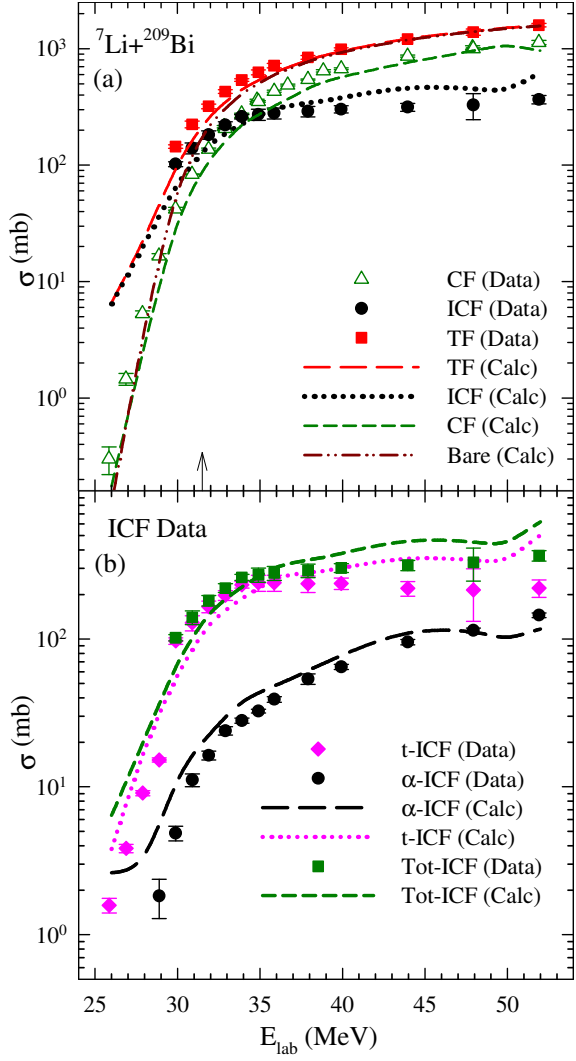}
\caption{\label{fig2} (Color online) Same as Fig. 1 but for $^{7}$Li+$^{209}$Bi reaction.}
\end{figure}

\begin{figure}
\includegraphics[width=0.44\textwidth,trim=0.3cm 7.5cm 10.5cm 1cm, clip=true]{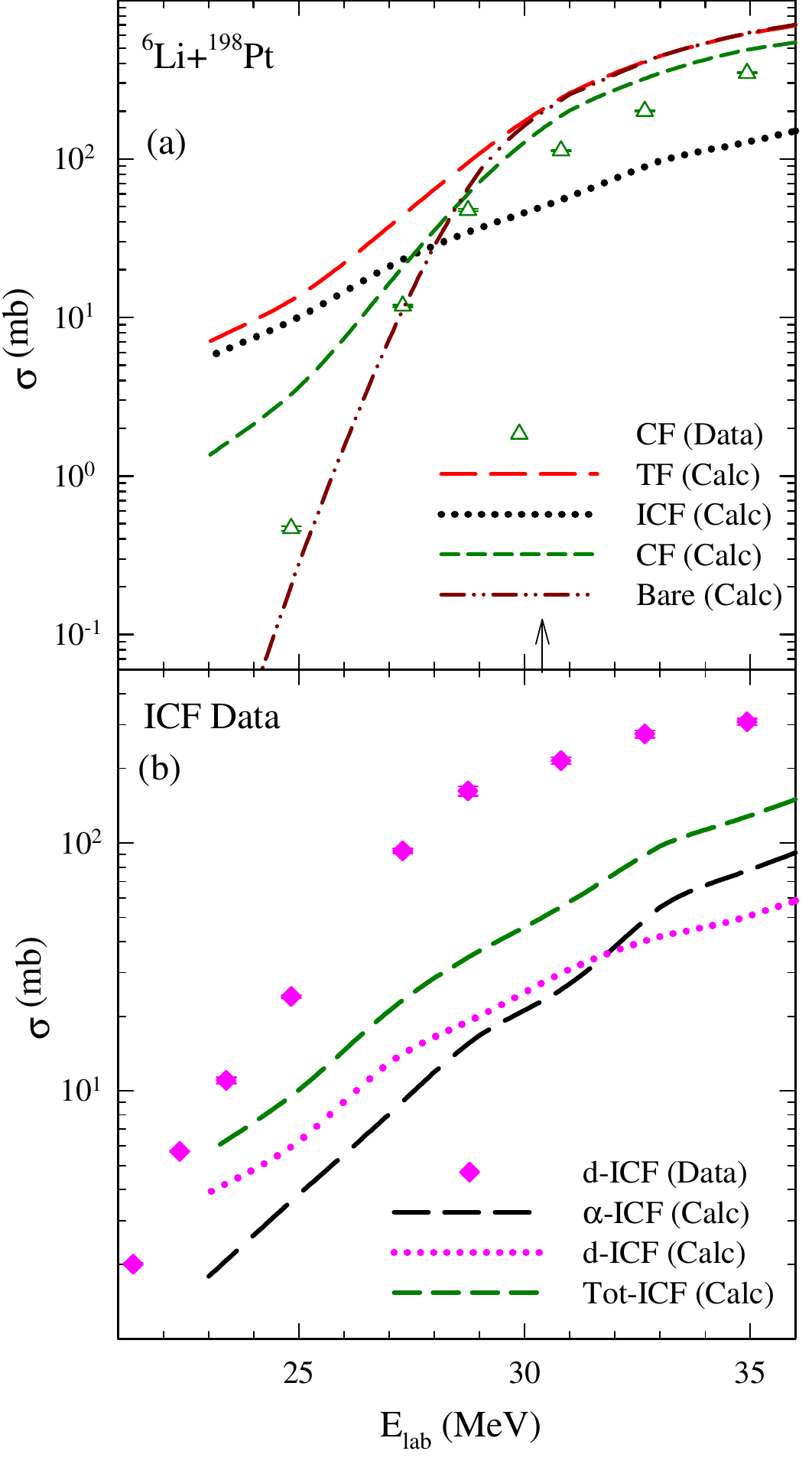}
\caption{\label{fig3} (Color online) Same as Fig. 1 but for $^{6}$Li+$^{198}$Pt reaction.}
\end{figure}

\begin{figure}
\includegraphics[width=0.44\textwidth,trim=0.3cm 7.5cm 10.5cm 1cm, clip=true]{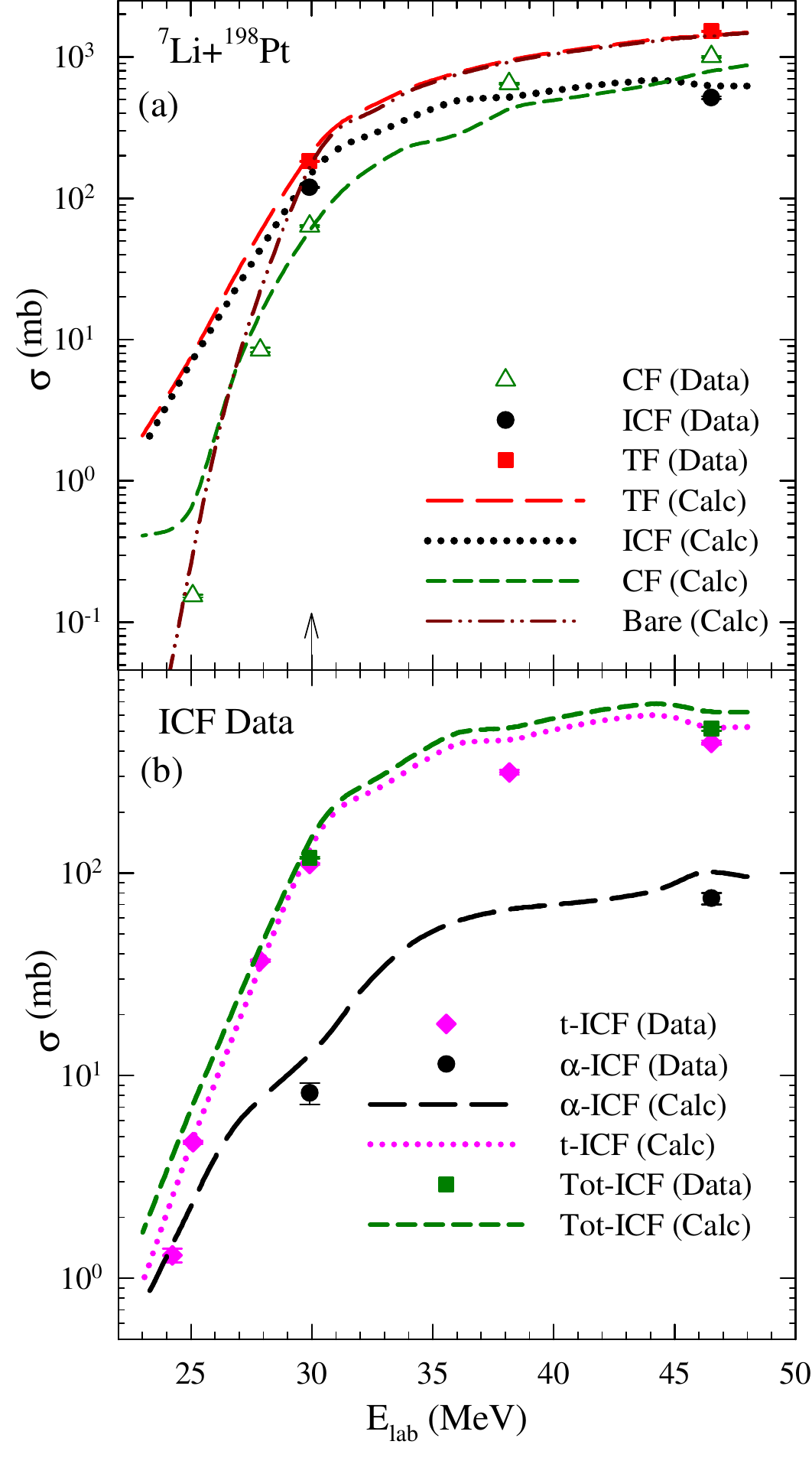}
\caption{\label{fig4} (Color online) Same as Fig. 1 but for $^{7}$Li+$^{198}$Pt reaction.}
\end{figure}

We have performed the detailed coupled channels calculations to study the fusion process for the $^{6,7}$Li+$^{209}$Bi and $^{6,7}$Li+$^{198}$Pt systems. The Continuum Discretized Coupled Channel (CDCC) calculations were performed using the code FRESCO version 2.9 \cite{Thomp88}. The coupling scheme used in CDCC is similar to that described in our earlier works \cite{Harphool08,Jha09}. In $^6$Li, couplings were included to the 1$^{+}$, 2$^{+}$, and 3$^{+}$ resonances and L = 0,1,2,3 $\alpha$-d continuum and in $^7$Li to {1/2}$^{-}$ first excited state, the {5/2}$^{-}$ and {7/2}$^{-}$ resonances and the L = 0,1,2,3 $\alpha$-t continuum. The binding potentials for $\alpha$-d in $^6$Li and $\alpha$-t in $^7$Li are taken from the Refs.\ \cite{Kubo72} and \cite{Buck88} respectively.

In the CDCC calculations, the fusion  cross sections can be obtained as the total absorption cross section, which is equal to the difference of the total reaction cross section $\sigma_R$ and the cross section of all explicitly coupled direct reaction channels $\sigma_{D}$. The absorption cross sections that are taken as the total fusion cross sections are in turn obtained from the  S-matrix elements given by,
\begin{equation} \nonumber
\sigma_{abs}= \sigma_{R} - \sigma_{D}
\label{eq:1}
\end{equation}
where \\
\begin{equation} \nonumber
\sigma_{R} = \frac{\pi}{k^2}\sum_{l}(2l+1)(1-|S_l|^2)
\end{equation}
$\hbar k$ represents the relative momentum of the two nuclei in the entrance channel. The required fragment-target potentials were generated in the cluster folding (CF) model using real potentials, $\it{viz.}$, V$_{\alpha-T}$  taken as Sau-Paulo potential \cite{Sau-Paulo}, while V$_{d-T}$ and V$_{t-T}$ were taken from Refs.\ \cite{Daehnick80} and \cite{Beccheeti69}, respectively. In the calculations presented here, the fusion cross-sections are first calculated by including the short-range imaginary (W$_{SR}$) volume type potentials in the coordinates of both projectile fragments relative to the target, as in Ref.\ \cite{Diaz03}. The fusion cross-section is calculated in terms of that amount of flux which leaves the coupled channels set (total absorption cross-section) because of the short-range imaginary part of the optical potential used for the fragment-target potentials. The use of this short-range imaginary potential is equivalent to the use of an incoming boundary condition inside the Coulomb barrier. The $\sigma_{TF}$ calculated in this way corresponds to a sum of the complete fusion cross section $\sigma_{CF}$ and two incomplete fusion cross sections, i.e., $\sigma_{\alpha-\textrm{ICF}}$ and $\sigma_{\textrm{d/t}-\textrm{ICF}}$. It is equivalent to the expectation value of the imaginary part with three body CDCC wave function of the system from both the bound and scattering states as discussed in Ref.\ \cite{Des15}.

The CDCC calculations with the breakup couplings are performed with three choices of optical potentials, where W$_{SR}$ is used for (i) both the projectile fragments relative to the target (Pot. A), (ii) the $\alpha$-T part only (Pot. B), and (iii) the d(t)-T part only (Pot. C). However, in all these calculations, an additional imaginary volume type potential with parameters W=25 MeV, r$_w$=1.0 fm and a$_w$=0.4 fm, without any real part is also present in the center of mass of the whole projectile for the projectile-target radial motion. With these potential choices, we perform three independent calculations one with all three imaginary components and other two where either of two fragment-target imaginary components are disabled. The cases where the imaginary part of a particular fragment is disabled means that the only one of the fragment is absorbed following the breakup and calculated absorption cross section reduces. Therefore, $\sigma_{TF}$ is reduced by the incomplete fusion of the other fragment. The differences allow us to make an estimate of individual ICF channel cross-sections. By performing these three independent calculations, one can evaluate all three quantities: (i) $\sigma_{TF}$, (ii) $\sigma_{\textrm{d/t}-\textrm{ICF}}$, and (iii) $\sigma_{\alpha-\textrm{ICF}}$ separately. The calculated CF cross-section is obtained from the subtraction of total ICF from TF cross-sections.

The optical model potentials for fragment-target interaction used in the CDCC calculations are given in Table\ \ref{Table1}. The present choice of optical potentials, where the real part has the standard parameters of the global potential and  the imaginary part has short-range character, successfully describes the measured elastic scattering data for the $^{6}$Li+$^{209}$Bi system providing a simultaneous description of the fusion and the elastic scattering data. A satisfactory agreement with the elastic scattering data is also obtained for $^{7}$Li+$^{208}$Pb elastic scattering angular distribution (since the data for the $^{7}$Li+$^{209}$Bi system is not available in the literature). In addition, the parameters of the short range imaginary potential in the range of r$_{w}$ = 0.6 to 1.0 fm and a$_{w}$ = 0.1 to 0.4 fm are found to be less sensitive for the calculation of $\sigma_{TF}$. However, in the calculation of ICF, the radius parameter of imaginary part is optimized with the higher energy $\alpha$-ICF and d\t-ICF data of $^{6,7}$Li+$^{209}$Bi systems and kept fixed for remaining energies and also for $^{6,7}$Li+$^{198}$Pt systems.

\begin{table}
\caption{\label{Table1} Optical model potentials for fragment-target interaction used in the CDCC calculations.}
\begin{tabular}
{cccccccc}
\hline
System & V$_{0}$ & r$_{0}$ & a$_{0}$ & W$_{SR}$ & r$_{w}$ & a$_{w}$ & \\
& (MeV) & (fm) & (fm) & (MeV) & (fm) & (fm) & \\
\hline \\
$\alpha$+$^{209}$Bi & Sau & Paulo & Pot & 25.0 & 0.62 & 0.40 &  \\
d+$^{209}$Bi & 96.5 & 0.97 & 0.74 & 25.0 & 0.66 & 0.40 &  \\
t+$^{209}$Bi & 160.8 & 0.97 & 0.72 & 25.0 & 0.69 & 0.40 &  \\
$\alpha$+$^{198}$Pt & Sau & Paulo & Pot & 25.0 & 0.62 & 0.40 &  \\
d+$^{198}$Pt & 97.7 & 0.96 & 0.74 & 25.0 & 0.66 & 0.40 &  \\
t+$^{198}$Pt & 161.3 & 0.96 & 0.72 & 25.0 & 0.69 & 0.40 &  \\
\hline \\
\end{tabular}
\label{tab1}
\end{table}

\section{\label{sec:Result} Results and Discussion}
\begin{figure}
\includegraphics[width=0.64\textwidth,trim=0.3cm 5.6cm 6.0cm 0.1cm, clip=true]{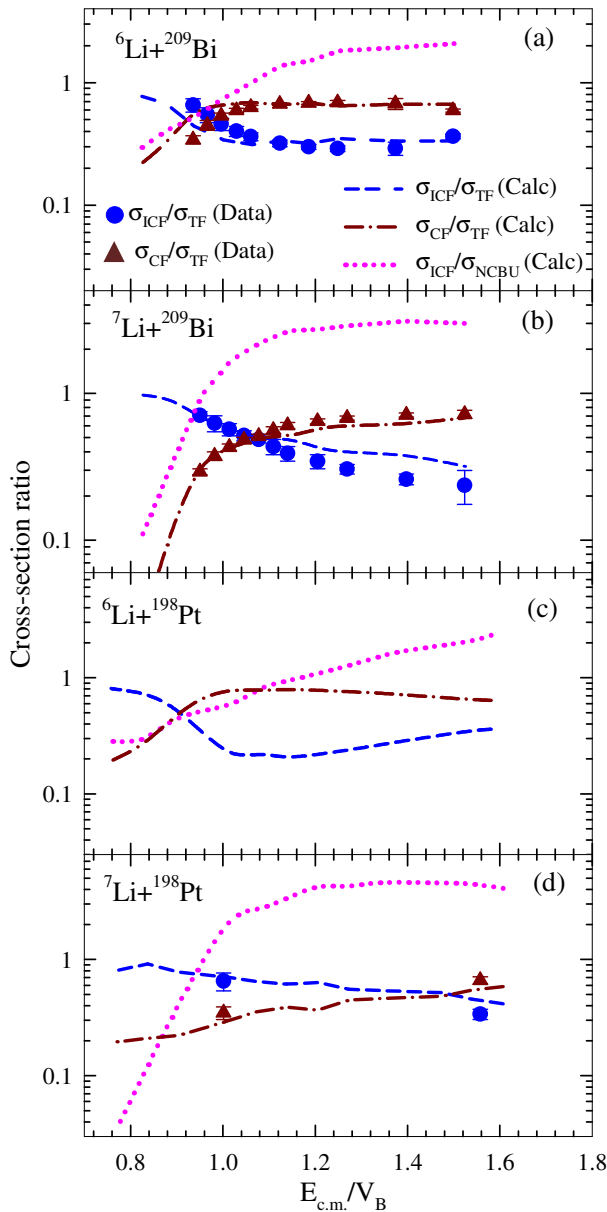}%
\caption{\label{ratiofig} (Color online) The ratio of cross-sections, $\sigma_{\textrm{ICF}}/\sigma_{\textrm{TF}}$, $\sigma_{\textrm{CF}}/\sigma_{\textrm{TF}}$ and $\sigma_{\textrm{ICF}}/\sigma_{\textrm{NCBU}}$ derived from the calculations as a function of E$_{\textrm{c.m.}}$/V$_{\textrm{B}}$ for $^{6,7}$Li+$^{209}$Bi and $^{6,7}$Li+$^{198}$Pt systems is shown by dashed, dashed-dot and dotted lines respectively. The symbols are showing the experimental data (see text for details).}
\end{figure}

In Figs.\ \ref{fig1}(a), \ref{fig2}(a), \ref{fig3}(a) and \ref{fig4}(a) results of the calculations for the  TF, CF and ICF cross-sections are shown with long dashed, short dashed and dotted lines, respectively along with the available measured data from Refs.\ \cite{Dasgupta04} and \cite{Shrivastava09,Shrivastava13} for $^{6,7}$Li+$^{209}$Bi and $^{6,7}$Li+$^{198}$Pt systems respectively. The bare calculations (without breakup couplings) were also performed and the calculated fusion cross-sections are denoted by dashed-dot-dot lines in the above mentioned figures. The Coulomb barrier positions are marked by arrow in all the figures. It is seen that at energies above the Coulomb barrier, the calculations which include the  couplings and calculations that omit them have negligible difference but at energies below the barrier, the coupled TF cross-sections are enhanced in comparison to bare TF cross-sections.

The individual ICF cross-sections, $\sigma_{\alpha-\textrm{ICF}}$ and $\sigma_{\textrm{d/t}-\textrm{ICF}}$, described in previous section are extracted and shown in Figs.\ \ref{fig1}(b), \ref{fig2}(b), \ref{fig3}(b) and \ref{fig4}(b). In these figures, the long dashed, dotted and short dashed lines are the $\alpha$-ICF, d/t-ICF and Tot-ICF calculations, respectively. As can be seen from Figs.\ \ref{fig1} (b) and \ref{fig2} (b), there is a better agreement of $\alpha$-ICF data with the calculations at all the energies. In the case of d-ICF data (Fig.\ \ref{fig1} (b)), the high energy d-ICF data is over predicted by the calculations. Similarly, for t-ICF data (Fig.\ \ref{fig2} (b)), while the overall agreement of calculations with the data is satisfactory, the high energy data is below the calculations. This may be because of the fact that the long lived residue from d or t-ICF followed by neutron evaporation path, i.e. $^{209}$Po was not measured in the experiment, which has a significant contribution in d/t ICF.

There is also an interesting observation that the cross-section of d-ICF and $\alpha$-ICF in $^{6}$Li+$^{209}$Bi is of similar order. However, in the case of $^{7}$Li+$^{209}$Bi, t-ICF cross-section is much higher than $\alpha$-ICF, which is also evident from the data. Intuitively, we expect that there should be larger d-ICF than the $\alpha$-ICF cross-section, as the Coulomb barriers seen by d+$^{209}$Bi is lower than that of $\alpha$+$^{209}$Bi. This logic holds true for t-ICF and $\alpha$-ICF as well in the $^{7}$Li+$^{209}$Bi reaction. In the case of $^{6}$Li+$^{198}$Pt system, higher energy data is not available unlike other system under study and also only d-ICF data (Fig.\ \ref{fig3} (b)) was available for comparison with the calculations. This data was found to be highly under predicted  by the calculations. Also, it was noticed that the d-ICF cross-section reported \cite {Shrivastava09} is even higher than the calculated total fusion cross-section at the last two measured energies, which is surprising. Besides this, for the $^{6}$Li+$^{198}$Pt system, the available d-capture data may also contain the contribution from the direct d-transfer to the target apart from the d-fusion component. Since, the calculations presented here only predict the d-fusion, it can be one of the reasons for the discrepancy apart from some possible uncertainties in the data measurements for this particular system. For $^{7}$Li+$^{198}$Pt system, both t-ICF and $\alpha$-ICF data is nicely explained by calculations (Fig.\ \ref{fig4} (b)). In this system also, we observe that the t-ICF cross-section is higher than $\alpha$-ICF, which is also evident from the data. For further confirmation of the above mentioned observations, we need more data of d-ICF, $\alpha$-ICF and t-ICF simultaneously in various systems like $^{6,7}$Li+$^{209}$Bi.

The measured d/t-ICF (alpha as outgoing channel) can also have the contributions from transfer induced breakup, which we have not taken into account in our calculations. Although, the transfer process is important, it is not found to significantly contribute to the measured inclusive alpha cross-sections for several systems. In the studies with $^{6}$Li on $^{209}$Bi \cite{Santra12} and $^{159}$Tb \cite{Pradhan13}, it is pointed out that the d-ICF cross-sections are much more dominant compared with the transfer cross-sections in inclusive alpha measurements. In recent complete study on transfer induced breakup with $^{7}$Li on $^{89}$Y \cite{Sanat16}, it is concluded that the transfer induced breakup and non-capture breakup added together can only account for 8 \% of inclusive alpha cross-section.

The ICF cross-sections calculated in the present work represents the absorption from the breakup channel and hence, it is expected that the non-capture breakup and ICF are competing processes. The ratio of cross-sections, $\sigma_{\textrm{ICF}}/\sigma_{\textrm{TF}}$, $\sigma_{\textrm{CF}}/\sigma_{\textrm{TF}}$ and $\sigma_{\textrm{ICF}}/\sigma_{\textrm{NCBU}}$ derived from the calculations as a function of E$_{\textrm{c.m.}}$/V$_{\textrm{B}}$ for $^{6,7}$Li+$^{209}$Bi and $^{6,7}$Li+$^{198}$Pt systems is shown by dashed, dashed-dot and dotted lines in Fig.\ \ref{ratiofig} respectively. The corresponding experimental data of $\sigma_{\textrm{ICF}}/\sigma_{\textrm{TF}}$ and $\sigma_{\textrm{CF}}/\sigma_{\textrm{TF}}$ is shown with filled circle and filled triangle respectively in Fig.\ \ref{ratiofig}. Following two observations can be made from the plots: (i)$\sigma_{\textrm{ICF}}/\sigma_{\textrm{TF}}$ and $\sigma_{\textrm{CF}}/\sigma_{\textrm{TF}}$ ratio remain approximately constant over the energy range above the Coulomb barrier. For energies below the barrier, the $\sigma_{\textrm{ICF}}/\sigma_{\textrm{TF}}$ ratio is increasing while $\sigma_{\textrm{CF}}/\sigma_{\textrm{TF}}$ ratio is decreasing. This shows the dominance of ICF at below barrier energies in TF over CF cross-sections. The $\sigma_{\textrm{ICF}}/\sigma_{\textrm{TF}}$ ratio at above barrier energies gives the value of suppression factor in CF, which is found to be in agreement ($\sim$ 30 \%) with the literature data with $^{6,7}$Li projectiles from various measurements \cite{Parkar10,Harphool12}. That means the suppression observed in CF at above barrier energies is commensurate with the value of ICF. (ii) $\sigma_{\textrm{ICF}}/\sigma_{\textrm{NCBU}}$ ratio is nearly constant at above barrier while it decreases at below barrier energies. This indicates that at below barrier, the probability of capturing one fragment from breakup (ICF) is much less than escape of both the fragments (NCBU), while at above barrier energies ICF gradually becomes more significant.

\section{\label{sec:Sum} Summary}
In the study of weakly bound nuclei, because of low breakup thresholds, the breakup channel plays an important role in reaction mechanism. Here we have attempted to find its effect on fusion cross-sections via Continuum Discretized Coupled Channel calculations. The cluster folding potentials in the real part along with short-range imaginary part were used to calculate the CF, ICF and TF cross-sections for $^{6,7}$Li+$^{209}$Bi and $^{6,7}$Li+$^{198}$Pt systems. The simultaneous explanation of the measured experimental data for the CF, ICF and TF cross-sections  over the entire energy range is obtained for the first time using calculations in the full quantum mechanical approach. The calculated TF cross-sections from uncoupled and coupled were found to match at energies above the barrier, while below barrier uncoupled TF is lower than the coupled one. The large difference between these results at below barrier energies, imply the strong role of breakup couplings at below barrier energies. The individual ICF cross-sections imply that the d-ICF cross-section is of similar order that of $\alpha$-ICF cross-section in case of $^{6}$Li, while t-ICF cross-section is much more than $\alpha$-ICF cross-section in case of $^{7}$Li, which is also evident from the experimental data. More data of d-ICF, $\alpha$-ICF and t-ICF simultaneously for various systems is required to further emphasize this observation. The calculated ICF fraction which is the ratio of ICF and TF as a function of energy is found to be constant at energies above the barrier and it increases at energies below the barrier. This ratio which signifies the suppression of CF in TF is constant at above barrier energies and it is in agreement with the available data for several systems. At below barrier, as the ratio increases, it shows the enhanced importance of ICF contribution in TF at below barrier energies. The ratio of calculated ICF to NCBU shows the greater importance of NCBU at below barrier energies, while at above barrier energies ICF gradually becomes more significant.

\begin{acknowledgments}
One of the authors (V.V.P.) acknowledges the financial support through the INSPIRE Faculty Program, from the Department of Science and Technology, Government of India, in carrying out these investigations.
\end{acknowledgments}

\end{document}